\documentclass[aps,prb,twocolumn,showpacs,superscriptaddress]{revtex4}
\usepackage{here}
\usepackage{graphicx}
\makeatletter
\def\ScaleIfNeeded{%
\ifdim\Gin@nat@width>\linewidth \linewidth \else \Gin@nat@width \fi
} \makeatother
\begin{document}
% Use the \preprint command to place your local institutional report
% number in the upper righthand corner of the title page in preprint mode.
% Multiple \preprint commands are allowed.
% Use the 'preprintnumbers' class option to override journal defaults
% to display numbers if necessary
%\preprint{}

\title{Magnetic Order and Spin Dynamics in the Proximity of a Ferromagnetic Quantum Critical Point: a $\mu$SR study of
YbNi$_{4}$P$_{2}$}

\author{J.~Spehling}
 \affiliation{Institute for Solid State Physics, TU Dresden, D-01069 Dresden, Germany}
\author{M.~G\"unther}
 \affiliation{Institute for Solid State Physics, TU Dresden, D-01069 Dresden, Germany}
\author{C.~Krellner}
 \affiliation{Max Planck Institute for Chemical Physics of Solids, D-01187 Dresden, Germany}
 \affiliation{Cavendish Laboratory, University of Cambridge, J~J~Thomson Avenue, Cambridge CB3 0HE, United Kingdom}
\author{N.~Y$\grave{e}$che}
 \affiliation{Institute for Solid State Physics, TU Dresden, D-01069 Dresden, Germany}
\author{H.~Luetkens}
 \affiliation{Laboratory for Muon-Spin Spectroscopy, Paul-Scherrer-Institute, CH-5232 Villigen, Switzerland}
\author{C.~Baines}
\affiliation{Laboratory for Muon-Spin Spectroscopy, Paul-Scherrer-Institute, CH-5232 Villigen, Switzerland}
\author{C.~Geibel}
\affiliation{Max Planck Institute for Chemical Physics of Solids, D-01187 Dresden, Germany}
\author{H.-H.~Klauss}
 \email{h.klauss@physik.tu-dresden.de}
  \affiliation{Institute for Solid State Physics, TU Dresden, D-01069 Dresden, Germany}

\date{\today}

\begin{abstract}

The local 4$f$-electronic spin dynamics and magnetic order in
YbNi$_{4}$P$_{2}$ were studied by means of muon-spin relaxation
measurements. Zero-field muon-spin relaxation proves static magnetic
order with a strongly reduced ordered Yb$^{3+}$ moment of
$(2.5-4.6)\times10^{-2}\mu_{B}$, below $T_{C}=140$~mK. Above
$T_{C}$, the muon spin polarization $P(t,B)$ is dominated by
quasihomogeneous spin fluctuations and exhibits a time-field scaling
relation $P(t,B)=P(t/B^{\gamma})$, indicating cooperative critical
spin dynamics in the system. At $T=190$~mK, slightly above $T_{C}$,
$\gamma=0.81(5)$ suggesting time-scale invariant power-law behavior
for the dynamic electronic spin-spin autocorrelation function.

\end{abstract}

% insert suggested PACS numbers in braces on next line
% insert suggested keywords - APS authors don't need to do this
%\keywords{}

%\maketitle must follow title, authors, abstract, \pacs, and \keywords
\maketitle

%\section{Introduction}

Lanthanide-based heavy-fermion (HF) systems are suitable model
systems to study emergent phenomena at a quantum critical point
(QCP), where collective quantum fluctuations trigger the system
continuously from a magnetically ordered to a non-magnetic ground
state~\cite{FocusIssue2008,Mathur1998,Stewart2001,Park2006,Friedemann2009,Stockert2011}.
However, despite intense research, to the best of our knowledge, no
4$f$-based material is known with a continuous ferromagnetic (FM) to
paramagnetic (PM) quantum phase transition (QPT). The existence of
such a QPT is also controversially discussed from a theoretical
point of
view~\cite{Kirkpatrick2003,Conduit2009,Yamamoto2010,Peters2012,Green2012}.

Recently, Krellner \textsl{et al.} suggested that the HF metal
YbNi$_{4}$P$_{2}$ with a quasi-one-dimensional (1-D) electronic
structure exhibits FM quantum criticality above a low FM transition
temperature $T_{C}=170$ mK~\cite{Krellner2011}. YbNi$_{4}$P$_{2}$
crystallizes in the tetragonal ZrFe$_{4}$Si$_{2}$ structure
containing isolated chains of edge-connected Ni tetrahedra along the
$c-$axis. The Yb atoms are located in the channels between these Ni
tetrahedral chains. The reduced dimensionality in the Yb and Ni
network and the geometrical frustration between neighboring Yb
chains give rise to enhanced quantum spin fluctuations of the
magnetic Yb$^{3+}$ ions. In the PM state above 50~K, the magnetic
susceptibility shows Curie-Weiss behavior with an effective moment
$\mu_{eff}=4.52\mu_{B}$ that is characteristic for magnetic
Yb$^{3+}$ ions. Analysis of the magnetic entropy reveals a Kondo
energy scale of $T_{K}\approx8$ K for the crystal electric field
ground state doublet. The FM transition is evidenced by distinct
anomalies in magnetic susceptibility, specific heat, and resistivity
measurements.  Low-$T$ magnetization measurements suggest an ordered
FM moment of $m_{ord}\approx0.05(4)\mu_{B}$. Pronounced
non-Fermi-liquid (NFL) behavior is reflected by a
stronger-than-logarithmic diverging Sommerfeld coefficient and a
linear-in-$T$ resistivity state apparent in a $T$ range larger than
a decade above $T_{C}$. In external magnetic fields, the NFL
behavior is suppressed and FL behavior gradually recovers.
Therefore, YbNi$_{4}$P$_{2}$ is considered as a clean system
situated in the very close vicinity of a FM QCP, with FM quantum
fluctuations dominating thermodynamic and transport quantities at
$T>T_{C}$.

The present knowledge on YbNi$_{4}$P$_{2}$ is based on measurements
of macroscopic magnetic, thermodynamic, and transport properties.
The next step in a deeper investigation of this prospective FM
quantum critical system is to get insight on a microscopic level.
Beside the nature of the magnetic order, a central issue in the
present context of critical behavior is the spin dynamics. Since in
systems close to a QCP, the ordered moment is usually strongly
reduced, muon spin relaxation ($\mu$SR) has proven to be an
extremely valuable technique to collect appropriate
information~\cite{MacLaughlin2001,Ishida2003,MacLaughlin2004}.

Here, we present $\mu$SR experiments on polycrystalline
YbNi$_{4}$P$_{2}$, providing microscopic evidence for static
magnetism at $T\leq T_{C}\approx140$ mK with an ordered moment of
$m_{ord}=(2.5-4.6)\times10^{-2}\mu_{B}$/Yb, depending on the assumed
muon site. Above $T_{C}$, the muon-spin polarization $P(t)$ obeys
the time-field scaling relation $P(t)=P(t/B^{0.81(5)})$, indicating
cooperative and critical spin dynamics.

In a $\mu$SR experiment positive spin-polarized muons are implanted
into the sample, and the subsequent time evolution of the muon spin
polarization is monitored by detecting the asymmetric spatial
distribution of positrons emitted from the muon decay~\cite{Schenk}.
$\mu$SR in longitudinal applied magnetic fields is dominated by
Yb-4$f$ electronic spin fluctuations that couple to the implanted
muons. The $\mu$SR experiments on YbNi$_{4}$P$_{2}$ in zero field
(ZF) and longitudinal (LF) applied field -- with respect to the
initial muon spin polarization -- were performed on the $\pi$M3 beam
line at the Swiss Muon Source (S$\mu$S) at the
Paul-Scherrer-Institut, Switzerland. The sample was prepared by
crushing $\sim270$~mg of single crystalline material, grown in a
self-flux at 1400$\,^{\circ}\mathrm{C}$ in a closed Tantal crucible
and characterized by powder x-ray diffraction experiments, proving
the absence of any foreign phases. Detailed low-$T$ measurements on
polycrystalline YbNi$_{4}$P$_{2}$ were reported
elsewhere~\cite{Krellner2011}.

%\section{Results and Discussion}

Figure~1(a) displays typical time dependencies of the ZF muon-spin
polarization $P(t)$ in YbNi$_{4}$P$_{2}$ at representative
temperatures above and below $T_{C}$. A finite $T$-independent
background signal due to muons that stopped in a Ag sample holder
(signal fraction $\approx50\%$) was taken into account. At
$T\geq160$~mK, an exponential muon-spin relaxation is associated
with fast fluctuating paramagnetic electron spins with a relaxation
rate $\lambda$(160~mK)$ \approx0.152(2)\mu s^{-1}$. Note, that dense
static nuclear dipole moments would give rise to a weak Gaussian
relaxation in the PM regime. While cooling through $T_{C}$, an
additional magnetic relaxation mechanism is apparent, strongly
increasing with lowering $T$. Below $T_{C}$, a low-frequency
oscillation with a Gaussian relaxation of the muon-spin polarization
is observed indicating magnetic ordering of weak electronic
Yb$^{3+}$ moments. The muon-spin asymmetry data in the FM regime can
be described best using the functional
form~\cite{Barsov1990,Kornilov1991}:

\begin{figure}
\begin{center}
\includegraphics[width=0.90\columnwidth]{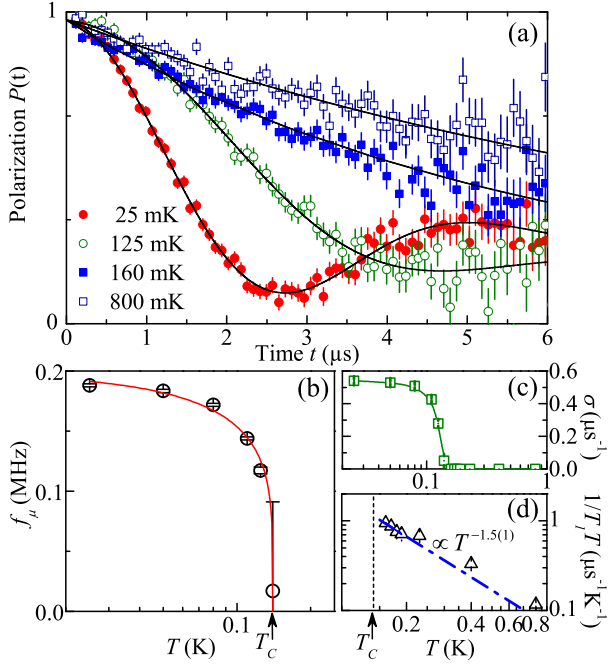}
\end{center}
\caption[1]{(Color online)~(a) Corrected muon spin polarization
$P(t)$ at ZF for representative $T$ above and below
$T_{C}\approx140$~mK. At $T\leq T_{C}$, solid lines are fitting
curves according to Eq.~(1). (b) $T$ dependence of the ZF $\mu$SR
frequency $f_{\mu}(T)$. The solid line is a fit to the
phenomenological function:
$f_{\mu}=f_{\mu}(0)\cdot(1-\frac{T}{T_{C}})^{n}$. (c) $T$ dependence
of the ZF static internal field distribution $\sigma$ in Eq.~(1).
The solid line is a guide to the eye. (d) $T$-dependence of
$1/T_{1}T$ in the PM regime. The line describes power-law behavior
as $\frac{1}{T_{1}T}\propto T^{-1.5}$.}
\end{figure}

\begin{eqnarray}
P(t)=\frac{1}{3}+\frac{2}{3}[\cos(2\pi
f_{\mu}t)-\frac{\sigma^{2}t}{2\pi f_{\mu}}\cdot\sin(2\pi
f_{\mu}t)]\cdot e^{-\frac{1}{2}\sigma^{2}t^{2}},
\end{eqnarray}

\begin{figure}
\begin{center}
\includegraphics[width=0.90\columnwidth]{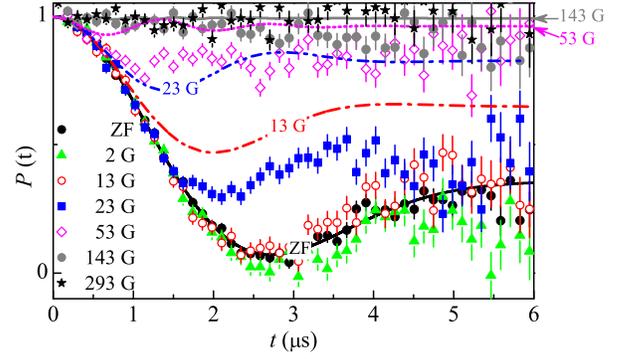}
\end{center}
\caption[1]{(Color online) Corrected muon spin polarization at
$T=20$~mK and various longitudinal magnetic fields $B_{LF}$. The
lines represent theoretical depolarization curves for the static GKT
function in corresponding longitudinal fields.}
\end{figure}

where $f_{\mu}$ and $\sigma$ are the muon spin precession frequency
and the Gaussian field width, respectively. The 2/3 oscillating and
the 1/3 non-oscillating terms originate from the spatial averaging
in polycrystalline samples, where 2/3 (1/3) of the internal magnetic
field components are directed perpendicular (parallel) to the
initial muon spin, causing a precession (no precession) of the muon
spin. The observation of a 2/3 and 1/3 signal fraction below
$T_{C}$, implies dense magnetic moments and proof that 100\% of the
sample volume shows static magnetic order. The latter is supported
by LF-$\mu$SR measurements as discussed in detail below. In the
limit $2\pi f_{\mu}\gg\sigma$, Eq.~(1) becomes a Gaussian damped
cosine function. For $2\pi f_{\mu}\rightarrow0$, close to the
magnetic transition, Eq.~(1) is equivalent to the Gaussian
Kubo-Toyabe (GKT) function~\cite{Hayano}, which describes a
muon-spin relaxation due to a static Gaussian field distribution
centered around $B_{local}=0$. ZF-$\mu$SR on the
antiferromagnetically ordered system
YbRh$_{2}$Si$_{2}$~\cite{Ishida2003} reveals a similar crossover
from a Lorentzian to a Gaussian damped $\mu$SR signal in the
vicinity of the PM to magnetic phase transition, attributed to a
transition from dynamic to static magnetism of magnetic Yb$^{3+}$
moments.

For YbNi$_{4}$P$_{2}$ a finite $\mu$SR frequency is clearly observed
below 150~mK. From the measured frequency value
$f_{\mu}=0.188(1)$~MHz at 20~mK one can determine the internal local
field at the muon site to $B_{local}=13.87$~G using $B_{local}=2\pi
f_{\mu}/\gamma_{\mu}$ with $\gamma_{\mu}=2\pi\times13.55$~kHz/G as
the muon gyromagnetic ratio. The local field $B_{local}$ as well as
the local static field width $\Delta
B_{local}=\sigma/\gamma_{\mu}\approx6$~G are very small for
conventional rare-earth magnets with large ordered moments. The
fractional width $\Delta B_{local}/B_{local}$ of the spontaneous
field distribution is $\sim0.4$ at low $T$ and remains constant as
$T\rightarrow T_{C}$, which is a reasonable value for a magnetically
ordered HF system, as e.g. in CeRhIn$_{5}$~$\Delta
B_{local}/B_{local} = 0.5$ is observed~\cite{Schenk2002}. Thus, the
local field distribution is nearly uniform and homogeneous in the FM
regime. The spontaneous muon-spin precession and the Gaussian shape
of the internal field distribution below $T_{C}$ arise from a dense
system of weak magnetic moments with small, static magnetic
inhomogeneities. The presence of a finite $B_{local}\neq0$ proves
coherent magnetic order.

ZF-$\mu$SR allows a precise determination of the $T$ dependence of
the magnetic order parameter, which is proportional to the measured
$\mu$SR frequency $f_{\mu}$. The $T$ dependence of $f_{\mu}$ and
$\sigma$ is shown in Figs.~1(b) and 1(c). For $T\leq140$~mK, both
observables exhibit a continuous increase. The $T$ dependence of
$f_{\mu}$ can be fit to the phenomenological function
$f_{\mu}=f_{\mu}(0)\cdot(1-\frac{T}{T_{C}})^{n}$ for $T<T_{C}$ with
$n=0.208\pm0.02$, $f_{\mu}(T)=0.199(3)$~MHz, and $T_{C}=140(2)$~mK.
The value of the effective critical exponent $n$, describing the
critical behavior close to $T_{C}$, is between $n=0.125$ and 0.325,
which are theoretically expected for two-dimensional (2D) and
isotropic three-dimensional (3D) Ising magnets, respectively. This
is not in contradiction with the claim of a quasi-1D system. In such
a system, the weak inter-chain coupling results in an evolution from
a 1D behavior at high $T$ to a 2D Ising or 3D behavior at low $T$,
which is intimately linked with (and is a prerequisite for) the
long-range ordering at finite $T$. The low data point density
between $0.6\leq\frac{T}{T_{C}}\leq1$, however, precludes the
determination of the precise critical exponent. The obtained value
for $T_{C}$ agrees well with the value found in specific heat
measurements on these single crystals~\cite{Steppke2011}.

\begin{figure}
\begin{center}
\includegraphics[width=0.90\columnwidth]{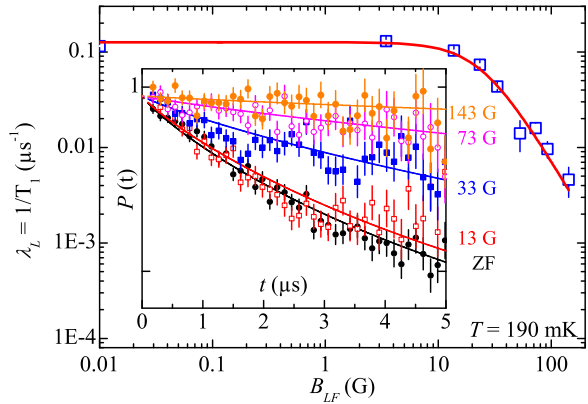}
\end{center}
\caption[1]{(Color online) Main panel: Field dependence of the
dynamic muon spin relaxation rate $\lambda_{L}$. The solid curve
represents a Redfield fit. For display reasons, the ZF value is set
at $B_{LF}=0.01$~G. Inset: Field dependence of the corrected muon
spin polarization $P(t)$ at $T=190$~mK.}
\end{figure}

For all examined $T\leq T_{C}$, the sample signal is analyzed with a
well-defined single $f_{\mu}$ and $\sigma$, signaling that the
magnetic order is a bulk effect and that only one dominant muon
stopping site is present. In general, for the determination of the
muon stopping site(s) it is important to deduce the hyperfine
coupling constant. One way to find potential muon sites is to
compare calculated and measured quantities for the local field
$B_{local}$ at the muon site. The muon preferentially settles at
tetrahedra or octahedra interstitial crystallographic sites. From
simple symmetry arguments the most probable muon stopping sites,
using Wyckoff's notation, are 4$f$(1/4,1/4,0), 8$j$(1/4,1/4,1/4),
4$f$(1/4,1/4,1/2), 8$i$(1/4,1/2,1/2), 4$c$(1/2,0,0),
4$c$(1/2,0,1/2), 2$b$(0,0,1/2), and 2$a$(1/2,1/2,1/2). For a
particular FM structure with the magnetic Yb$^{3+}$ moments aligned
within the $a~b$ plane and a dominant 4$f$-$\mu$ dipolar
interaction, one can determine the expected internal field values
for the proposed sites. Our lattice sum calculations reveal that
only at the 4$c$(1/2,0,1/2) site and 8$i$(1/4,1/2,1/2) site a local
field $B_{local}$ of the measured absolute magnitude is found. For
the 4$c$ and 8$i$ sites the measured local field of
$B_{local}=13.87$~G corresponds to a static ordered moment of the Yb
ions of $m_{ord}=0.046\mu_{B}$ and 0.025$\mu_{B}$, respectively.
Both values are in good agreement with the value deduced from recent
magnetization measurements~\cite{Krellner2011}.

\begin{figure}
\begin{center}
\includegraphics[width=0.90\columnwidth]{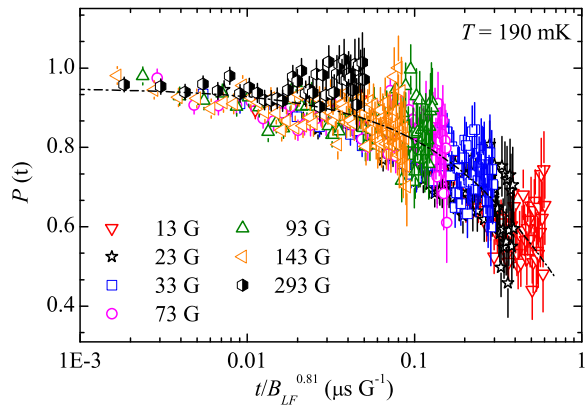}
\end{center}
\caption[1]{(Color online) Corrected muon decay asymmetry at
$T=190$~mK for various magnetic fields as function of the scaling
variable $t/B_{LF}^{0.81}$. The dashed-dotted line is a fit of the
13~G data with $P(t)/P(0)=\exp[-\lambda_{L}t]^{-0.9}$.}
\end{figure}

The temperature dependence of the exponential relaxation rate
$\lambda_{L}=\frac{1}{T_{1}}$, observed above $T_{C}$, is plotted in
Fig.~1(d) on a log-log scale as $\frac{1}{T_{1}T}$. Cooling down
from 800~mK, $\frac{1}{T_{1}T}$ exhibits power-law behavior
according to $\frac{1}{T_{1}T}\propto T^{-1.40(6)}$. At
$T\leq190$~mK, the power-law behavior in $\frac{1}{T_{1}T}$ persists
in the PM regime down to $T_{C}$, however, with a slight change of
the critical exponent, i.e., $\frac{1}{T_{1}T}\propto T^{-1.5(1)}$
(dashed-dotted line). The observed $\frac{1}{T_{1}T}$ behavior is
close to the $T^{-4/3}$ temperature dependence predicted by the
self-consistent renormalization (SCR) theory for a system close to a
3D ferromagnetic QCP~\cite{Moriya1995}. There is no prediction for
an itinerant quasi 1D system in the $T$ range between the exchange
energy scale and ordering temperature. For an isolating
ferromagnetic quasi 1D spin chain the $T$ dependence of the
relaxation rate above $T_{C}$ depends strongly on the details of the
interactions -- see, e.g.,~\cite{Sato2011}.

LF-$\mu$SR experiments allow to separate the dynamic contribution to
the relaxation of the muon-spin polarization. Investigations of the
low-$T$ muon-spin dynamics yield additional information about the
origin of the NFL behavior in YbNi$_{4}$P$_{2}$. Figure~2 displays
the muon-spin asymmetry function $P(t)$ at $T=20$~mK for different
applied LF's. The muon-spin relaxation is completely suppressed in
an applied field $B_{LF}\approx300$~G, demonstrating that the
internal field distribution is static in nature. However, the
observed decoupling can not be described accurately by a standard
muon asymmetry function that considers an internal field
distribution which is symmetric around $B_{local}=0$. For
comparison, Fig~2. shows theoretical depolarization curves for the
static GKT function in the corresponding longitudinal magnetic
fields. This supports the ZF data, i.e., the observation of a broad
field distribution centered around a finite but small internal field
$B_{local}$(20~mK) $\approx13.87$~G in the FM phase. Finally, when
$B_{LF}\gg B_{local}$, the muon spin relaxation is decoupled from
the static $B_{local}$ as observed for $B_{LF}\geq23$~G.

At $T>T_{C}$, the field dependence of the muon-spin relaxation
probes the Fourier transform of the dynamic spin-spin
autocorrelation function
$q(t)=\langle\textbf{S}_{i}(t)\cdot\textbf{S}_{i}(0)\rangle$, which
exhibits exponential behavior for homogeneous systems and power-law
(or cutoff power-law) or stretched exponential behavior for
inhomogeneous systems. The inset of Fig.~3 displays the muon-spin
polarization $P(t)$ at $T=190$~mK, both in magnetic LF between 13
and 143~G and ZF. The relaxation rate $\lambda_{L}$ is reduced with
increasing field. The field dependence of $\lambda_{L}$ is given in
the main panel of Fig.~3. It shows nearly no field dependence for
magnetic fields of less than $\sim13$~G, but varies more strongly,
as $H^{-\kappa}$ with $\kappa\approx0.79(7)$, for higher fields.
From the field dependence of $\lambda_{L}$, the spin autocorrelation
time $\tau_{c}$ can be estimated using the Redfield formalism for
$\lambda_{L}(B_{LF})=(2\gamma_{\mu}^{2}\langle
B_{fluc}^{2}\rangle\tau_{c})/[1+(\gamma_{\mu}^{2}B_{LF}^{2}\tau_{c}^{2})]$
considering $\tau_{c}$ as independent of the applied field $B_{LF}$.
Here, $B_{fluc}(t)$ describes the time-varying local magnetic field
at the muon site due to fluctuations of neighboring Yb$^{3+}$
moments, with a local time averaged second moment
$\Delta^{2}=\gamma_{\mu}^{2}\langle B_{fluc}^{2}\rangle$ and a
single fluctuation time $\tau_{c}$. For $\hbar\omega\ll k_{B}T$
($\omega$ giving the spin fluctuation rate), the
fluctuation-dissipation theorem~\cite{Toll1956} relates $\tau_{c}$
to the imaginary component of the local $q$-independent $f$-electron
dynamic susceptibility, i.e.
$\tau_{c}(B)=(k_{B}T)[\chi^{\prime\prime}(\omega)/\omega]$. The fit
to the data (solid curve in the main panel of Fig.~3) yields
$\Delta^{2}\approx0.1$~(MHz) and $\tau_{c}\approx6\times10^{-7}$~s,
the latter value nearly three orders of magnitude larger than the
one obtained for YbRh$_{2}$Si$_{2}$ at $T=20$~mK~\cite{Ishida2003},
suggesting very slow critical fluctuations.

The $\mu$SR time spectra in Fig.~3 are well described with a
stretched exponential relaxation function of the form
$P(t)=P(0)\exp[-(\lambda t)^{\beta}]$. An exponent of
$\beta\approx0.9$ shows that the relaxation rate is nearly uniform
throughout the sample, indicating that YbNi$_{4}$P$_{2}$ exhibits
quasihomogeneous spin fluctuations for $T\ll T_{K}$. The spin
dynamics is characterized by a narrow distribution of correlation
times ($\beta=1$ corresponds to one single correlation time). Thus,
disorder-driven theories, including Kondo
disorder~\cite{Miranda1996,Miranda} and the Griffith phase
scenario~\cite{Neto1998} as primary mechanisms for the observed NFL
behavior, can be ruled out. It further implies that the crystalline
disorder in YbNi$_{4}$P$_{2}$ is quite small, which is consistent
with a small residual resistivity ($\rho_{0}\sim2.4\mu\Omega~cm$)
and the stoichiometric occupation of the crystallographic lattice
sites revealed by the x-ray structure
refinement~\cite{Krellner2011}.

A sensitive test to identify power-law or stretched exponential
behavior of $q(t)$ is a time-field scaling analysis of the muon-spin
relaxation function. In both cases a specific time-field scaling can
be found, i.e., the muon-spin relaxation function $P(t,B_{LF})$
obeys the scaling relation $P(t,B_{LF})=P(t/B_{LF}^{\gamma})$. This
relation applies only in the asymptotic strong field limit, i.e., as
long as $2\pi
f_{\mu}=\gamma_{\mu}B_{LF}\gg\lambda_{L}$~\cite{Keren}. If
time-field scaling is obeyed, a plot of $P(t,B_{LF})$ versus
$t/B_{LF}^{\gamma}$ at $T>T_{C}$ will be universal for the correct
choice of $\gamma$, and distinguishes between power-law ($\gamma<1$)
and stretched exponential ($\gamma\geq1$) correlations. For small
$B_{LF}$, the field dependence is expected to be due to the change
of $f_{\mu}$ rather than an effect of field on $q(t)$. A breakdown
of time-field scaling would occur for high fields where $q(t)$ is
directly effected by the applied fields. Figure~4 shows the same
asymmetry data, as displayed in Fig.~3, as functions of the scaling
variable $t/B_{LF}^{\gamma}$. For $\gamma=0.81(5)$ the data scale
well over $\sim2.5$ orders of magnitude in $t/B_{LF}^{\gamma}$ and
for all fields between 13 and 143~G, except for 293~G. Here, at
large $t$, the data fall above the low-field scaling curve. Fields
$\mu_{B}B_{LF}\geq k_{B}T$ (with $k_{B}=$Boltzmann`s constant) would
be expected to affect the spin dynamics. The scaling exponent
$\gamma=0.81(5)<1$ implies that within the $\mu$SR frequency range,
the spin-spin correlation function $q(t)$ is approximated by a power
law (or a cutoff power law) rather than a stretched exponential or
exponential~\cite{Keren}, consistent with the Redfield analysis. The
power-law is time-scale invariant and dynamical modulations should
therefore be observable in any time window. The obtained time-field
scaling of the relaxation data is a signature of slow homogeneous
spin dynamics. It strongly indicates that the critical slowing down
of spin fluctuations at the magnetic phase transition occurs
cooperatively throughout the sample. In stoichiometric, homogeneous
NFL systems such behavior may arise from the effect of disorder on
quantum critical fluctuations inherent to a QCP. This is suggested
for the NFL compound
YbRh$_{2}$Si$_{2}$~\cite{Ishida2003,MacLaughlin2004}.

%\section{CONCLUSION}

In conclusion, ZF-$\mu$SR in the stoichiometric NFL compound
YbNi$_{4}$P$_{2}$ clearly proves static magnetic ordering of
strongly reduced Yb$^{3+}$ moments below $T_{C}=140$~mK. Above
$T_{C}$, the muon spin polarization $P(t)$ obeys the time-field
scaling relation $P(t)=P(t/B^{0.81(5)})$ for applied magnetic fields
$B$ between 13 and 143~G, indicating cooperative and critical spin
dynamics. Power-law behavior of the dynamic spin-spin
autocorrelation function is implied by the observation of
$\gamma<1$~\cite{Keren}. The LF-$\mu$SR results suggest that the NFL
behavior observed at $T>T_{C}$ is induced by quasi homogeneous
critical spin fluctuations.

%\section{Acknowledgments}

We acknowledge with thanks the help of A. Amato and the PSI
accelerator crew as well as financial support by the German Science
Foundation (DFG) in the framework of the priority program 1458,
Grant No. KL1086/10-1.


\begin{references}

\bibitem{FocusIssue2008} Focus issue on quantum phase transitions, Nat. Phys. \textbf{4}, 167 - 204 (2008).
\bibitem{Mathur1998} N.~D.~Mathur, F.~M.~Grosche, S.~R.~Julian, I.~R.~Walker, D.~M.~Freye, R.~K.~W.~Haselwimmer, and
G.~G.~Lonzarich, Nature (London) \textbf{394}, 39 (1998).
\bibitem{Stewart2001} G.~R.~Stewart, Rev. Mod. Phys. \textbf{73}, 797 (2001); \textbf{78}, 743
(2006).
\bibitem{Park2006} T.~Park, F.~Ronning, H.~Q.~Yuan, M.~B.~Salamon, R.~Movshovich, J.~L.~Sarrao,
and J.~D.~Thompson, Nature (London) \textbf{440}, 65-68 (2006).
\bibitem{Friedemann2009} S.~Friedemann, T.~Westerkamp, M.~Brando, N.~Oeschler, S.~Wirth, P.~Gegenwart, C.~Krellner, C.~Geibel,
and F.~Steglich, Nature Phys. \textbf{5}, 465-469 (2009).
\bibitem{Stockert2011} O.~Stockert, J.~Arndt, E.~Faulhaber, C.~Geibel, H.~S.~Jeevan, S.~Kirchner, M.~Loewenhaupt, K.~Schmalzl,
W.~Schmidt, Q.~Si, and F.~Steglich, Nature Phys. \textbf{7}, 119-124
(2011).
\bibitem{Kirkpatrick2003} T.~R.~Kirkpatrick and D.~Belitz, Phys. Rev.
B \textbf{67}, 024419 (2003).
\bibitem{Conduit2009} G.~J.~ Conduit, A.~G.~Green, and B.~D.~Simons,
Phys. Rev. Lett. \textbf{103}, 207201 (2009).
\bibitem{Yamamoto2010} S.~J.~Yamamoto and Q.~Si, Proc. Natl. Acad. Sci. USA, \textbf{107}, 15704-15707 (2010).
\bibitem{Green2012} U.~Karahasanovic, F.~Kr\"uger and
A.~G.~Green, Phys. Rev. B \textbf{85}, 165111 (2012).
\bibitem{Peters2012} R.~Peters, N.~Kawakami and T.~Pruschke, Phys. Rev. Lett. \textbf{108}, 086402 (2012).
\bibitem{Krellner2011} C.~Krellner, S.~Lausberg, A.~Steppke, M.~Brando, L.~Pedrero,
H.~Pfau, S.~Tenc$\grave{e}$, H.~Rosner, F.~Steglich, and C.~Geibel,
New J. Phys. \textbf{13}, 103014 (2011).
\bibitem{MacLaughlin2001} D.~E.~MacLaughlin, O.~O.~Bernal, R.~H.~Heffner, G.~J.~Nieuwenhuys, M.~S.~Rose, J.~E.~Sonier, B.~Andraka,
R.~Chau, and M.~B.~Maple, Phys. Rev. Lett. \textbf{87} 066402
(2001).
\bibitem{Ishida2003} K.~Ishida, D.~E.~MacLaughlin, Ben-Li~Young, K.~Okamoto, Y.~Kawasaki, Y.~Kitaoka, G.~J.~Nieuwenhuys,
R.~H.~Heffner, O.~O.~Bernal, W.~Higemoto, A.~Koda, R.~Kadono,
O.~Trovarelli, C.~Geibel, and F.~Steglich, Phys. Rev. B \textbf{68},
184401 (2003).
\bibitem{MacLaughlin2004} D.~E.~MacLaughlin, R.~H.~Heffner, O.~O.~Bernal, K.~Ishida, J.~E.~Sonier,
G.~J.~Nieuwenhuys, M.~B.~Maple and G.~R.~Stewart, J. Phys.: Condens.
Matter \textbf{16}, 4470-4498 (2004) and references therein.
\bibitem{Schenk} A.~Schenck, \textsl{Muon Spin Rotation Spectroscopy: Principles and Applications in Solid State Physics}
(A. Hilger, Bristol and Boston, 1985).
\bibitem{Barsov1990} S.~G.~Barsov \textsl{et al.}, Hyperfine Interactions \textbf{64}, 415 (1990).
\bibitem{Kornilov1991} E.~I.~Kornilov and V.~Yu.~Pomjakushin, Phys. Lett. A \textbf{153}, 364 (1991).
\bibitem{Hayano} R.~S.~Hayano, Y.~J.~Uemura, J.~Imazato, N.~Nishida, T.~Yamazaki, and R.~Kubo, Phys. Rev. B \textbf{20}, 850
(1979).
\bibitem{Schenk2002} A.~Schenck, D.~Andreica, F.~N.~Gygax, D.~Aoki and Y.~\={O}nuki Phys. Rev. B \textbf{66}, 144404 (2002).
\bibitem{Steppke2011} A.~Steppke, \textsl{private communication}.
\bibitem{Moriya1995} T.~Moriya and T.~Takimoto, J. Phys. Soc. Jpn \textbf{64},
960 (1995); A.~Ishigaki and T.~Moriya, J. Phys. Soc. Jpn
\textbf{65}, 3402 (1996); T.~Misawa, Y.~Yamaji, and M.~Imada, J.
Phys. Soc. Jpn. \textbf{78}, 084707 (2009).
\bibitem{Sato2011} M.~Sato, T.~Hikihara, and T.~Momoi, Phys. Rev. B \textbf{83}, 064405 (2011).
\bibitem{Toll1956} J.~S.~Toll, Phys. Rev. \textbf{104}, 1760 (1956).
\bibitem{Miranda1996} E.~Miranda, V.~Dobrosavljevic, and G.~Kotliar  J. Phys.:
Condens. Matter \textbf{8} 9871 (1996).
\bibitem{Miranda} E.~Miranda, V.~Dobrosavljevic, and G.~Kotliar,
Phys. Rev. Lett. \textbf{78}, 290 (1997).
\bibitem{Neto1998} A.~H.~Castro Neto, G.~Castilla, and B.~A.~Jones, Phys. Rev. Lett. \textbf{81}, 3531 (1998).
\bibitem{Keren} A.~Keren, P.~Mendels, I.~A.~Campbell, and J.~Lord, Phys. Rev. Lett. \textbf{77},
1386 (1996).


\end{references}
\end{document}